\documentstyle[psfig,floats,aps,rmp,harvard,eqsecnum]{revtex}

\citationstyle{dcu}
\begin{document}

\twocolumn[\hsize\textwidth\columnwidth\hsize\csname@twocolumnfalse\endcsname

\author {$^1$Peter B. Weichman, $^2$Alexa W. Harter, and $^2$David L. Goodstein  }
\address {$^1$Blackhawk Geometrics, 301 Commercial Road, 
Suite B, Golden, CO 80401 \\
$^2$Condensed Matter Physics 114-36, 
California Institute of Technology, Pasadena, CA 91125}

\title{Criticality and Superfluidity in Liquid $^4$He Under Nonequilibrium
Conditions}

\date {{\today}}

\maketitle

\begin{abstract}

We review a striking array of recent experiments, and their theoretical
interpretations, on the superfluid transition in $^4$He in the presence 
of a heat flux, $Q$.  We define and evaluate a new set of critical point 
exponents.  The statics and dynamics of the superfluid-normal interface are 
discussed, with special attention to the role of gravity. If $Q$ is in the 
same direction as gravity, a self-organized state can arise, in which the
entire sample has a uniform reduced temperature, on either the normal or
superfluid side of the transition. Finally, we review recent theory and
experiment regarding the heat capacity at constant $Q$.  The excitement
that surrounds this field arises from the fact that advanced thermometry
and the future availability of a microgravity experimental platform
aboard the International Space Station will soon open to experimental
exploration decades of reduced temperature that were previously
inaccessible.

\end{abstract}
%\pacs{PACS numbers 05.70.Jk, 05.70.Ln, 64.60.Ht, 67.40.Pm}
\vskip2pc]
\narrowtext

\tableofcontents

\section{Introduction}  

Due to its extraordinary purity and insensitivity to external
perturbations, superfluid $^4$He has long been the best system for
accurate, detailed experimental investigations of phase transitions and
critical phenomena.  The essence of superfluidity, however, lies in the
dynamics of flowing helium, rather than in equilibrium properties such
as the specific heat and the superfluid density.  In recent years,
considerable experimental attention has been focused on the behavior of
liquid helium, near the lambda transition temperature, $T_\lambda$, when
a flux of heat, ${\bf Q}$, is passed through it \cite{DAS88,LA96,soc97,unm98,cq99}. 
This article reviews the wealth of exciting new physical phenomena uncovered by 
these experiments and by the parallel theoretical investigations
\cite{O83,HD92,HD94,WPMM98,H99a,WM00}.

At sufficiently small ${\bf Q}$, superfluid $^4$He transports heat
essentially without dissipation, by means of superfluid-normal
fluid counterflow.  At higher ${\bf Q}$, or as $T$ approaches
$T_\lambda$, this transport mode breaks down.  In Sec.\ II we
discuss the phase diagram of helium in the $Q$-$T$ plane.  In this plane,
the superfluid state is bounded by a transition curve, $Q_c(T)$,
outside of which dissipative flow takes over.  We also introduce and 
evaluate a new set of critical exponents that arise as a consequence
of superflow.

In Sec.\ III we describe an inhomogeneous phase where the heat flows
through both normal and superfluid regions, separated by an interface 
region that is neither normal nor superfluid.  In the normal region the 
heat flux produces a static temperature gradient.  In the superfluid 
region, the heat flows at constant temperature.  In the interface region, 
a transition between these types of behavior is mediated by fluctuations
in a way that is not yet accessible either to experiment or to theory.

A recurring theme throughout the article is the fundamental limitation 
imposed by the Earth's acceleration due to gravity, $g_e$, on the resolution 
of Earth-based experiments.  Gravity produces a pressure gradient across 
the sample, leading to a variation in the local lambda transition temperature 
$T_\lambda(z)$ with height $z$ according to 
\begin{equation}
\partial_z T_\lambda(z) = - \gamma {g \over g_e},
~~\gamma \simeq 1.27~{\mu{\rm K} \over {\rm cm}}.
\label{1.1}
\end{equation}
For generality, the possibility of gravity $g$ different from that on 
Earth is included.  For $g=g_e$ the transition temperature therefore
decreases by $1.27\mu$K per centimeter of column height.  If $T$
is tuned so that, say, the center of the cell is at the local lambda 
point, the upper region will be superfluid while the lower region will 
be normal fluid.  The interface between, defined as the region over which 
the local properties differ significantly from those of a bulk system at the 
same local temperature and pressure, is about 0.1~mm wide on Earth \cite{MG89}.  
This figure also estimates the maximum possible critical correlation 
length in the system.  

The inhomogeneity induced by $g_e$ causes the critical singularities 
to be rounded on a scale set by the height of the cell.  Balancing gravity 
effects against finite size effects, which enter if the cell is too small, 
the optimal cell height is about 0.3~mm, leading to rounding on a 
scale $\epsilon \equiv (T-T_\lambda)/T_\lambda \sim 10^{-7}$.  Present 
thermometry \cite{lpe96,unm98} is capable of resolving reduced temperatures 
$\epsilon$ that are 3--4 orders of magnitude smaller than this.  In essence 
there is a new frontier, consisting of decades of previously unavailable reduced 
temperature around $T_\lambda$, that can only be explored in microgravity.  
Experiments measuring equilibrium specific heat (Lambda Point Experiment---LPE) 
\cite{lpe96} and finite size effects (Confined Helium Experiment---CHEX) have 
recently flown aboard the space shuttle.

Gravity also has a strong effect on the interface region 
discussed in Sec.\ III.  As we have seen above, the interface is
compressed by gravity to a width of about 0.1~mm, too small to be 
studied experimentally.  The interface region will be studied in
an experiment (Critical Dynamics Experiment---DYNAMX) presently 
being prepared for a low temperature microgravity platform (Low 
Temperature Microgravity Physics Facility---LTMPF) that is to be 
part of the International Space Station (ISS).

Although gravity is detrimental to some measurements,
there can also be interesting new physics when both gravity and 
heat current are present.  In Sec.\ IV we describe the so-called 
\emph{self-organized critical} (SOC) state, in which 
$g$ and $Q$ conspire to produce a new, essentially homogeneous, 
nonequilibrium state where the temperature $T(z)$ precisely parallels
$T_\lambda(z)$ at a fixed $Q$-dependent distance.  At larger $Q$,
this state undergoes a transition from normal to superfluid, with the 
temperature gradient supported by a stream of vortices in the latter.

Finally, in Sec.\ V we discuss the specific heats under heat flow,
predicted \cite{HD94,cq96,HD96}, and subsequently confirmed
\cite{cq99}, to be \emph{enhanced} above the equilibrium form, with
a singularity predicted at the phase boundary $Q_c(T)$.  This experiment 
too ultimately needs to be performed at very low $Q$, and the required 
microgravity version is currently being proposed.

\section{Nonequilibrium critical phenomena and scaling}

We begin by presenting the basic mathematical background within 
which the physical phenomena we discuss in later sections is most 
clearly described and understood.  The language of phase transitions 
and critical phenomena, under both equilibrium and nonequilibrium 
conditions, is that of critical exponents, scaling relations, and 
scaling functions, which quantify the scale invariant nature of
a system near its critical point.  In order to keep the discussion 
at as elementary a level as possible, we will introduce the theory 
in close analogy to that appropriate to equilibrium classical magnetic 
and liquid-vapor critical points.  The near-dissipationless nature of 
superflow in $^4$He makes this analogy especially fruitful in
describing nonequilibrium superfluidity.

\subsection{Superfluid counterflow}

Under conditions where a uniform heat flux ${\bf Q}$ is applied
to superfluid $^4$He, a thermal counterflow is created in which 
the normal fluid moves along with ${\bf Q}$ at velocity ${\bf u}_n$, 
and the superfluid moves oppositely with velocity 
${\bf u}_s$.  For sufficiently small heat current not too close
to $T_\lambda$ [more precisely, one requires $Q < Q_c(T)$, with
$Q_c(T)$ defined in Fig.~\ref{fig2} and (\ref{2.20}) below] the 
temperature $T < T_\lambda$ is essentially uniform and the heat 
flux is
\begin{equation}
{\bf Q} = T S \rho {\bf u}_n,
\label{2.1}
\end{equation}
where $S$ is the entropy per unit mass, $\rho = \rho_s + \rho_n$ is 
the total mass density, composed of superfluid and normal fluid parts, 
and ${\bf u}_n$ is the normal fluid velocity.  Since there is no net 
mass flow, one has ${\bf j}_n = - {\bf j}_s$, where ${\bf j}_s = 
\rho_s {\bf u}_s$ and ${\bf j}_n = \rho_n {\bf u}_n$ are, respectively, 
the superfluid and normal fluid mass current densities, and ${\bf u}_s$ 
is the superfluid velocity.  Close to $T_\lambda$ one finds experimentally 
$S \simeq S_\lambda \equiv 1.58$J/gK, $\rho_s \approx \rho_0 |\epsilon|^\zeta$, 
with $\rho_0 \simeq 0.37$g/cm$^3$, critical exponent $\zeta \simeq 0.671$, and 
$\rho_n \simeq \rho \simeq 0.14$g/cm$^3$.  Thus, in experimentally motivated 
units,
\begin{eqnarray}
u_s &\approx& 7.9 \times 10^{-3} {Q \over 1 \mu{\rm W}/{\rm cm}^2}
\left({10^{-6} \over |\epsilon|} \right)^\zeta {\rm cm}/{\rm s}
\label{2.2} \\
u_n &\approx& 2.1 \times 10^{-6} {Q \over 1 \mu{\rm W}/{\rm cm}^2}\
{\rm cm}/{\rm s}.
\label{2.3}
\end{eqnarray}
At experimentally accessible values $Q=1\mu$W/cm$^2$ and $|\epsilon| 
= 10^{-6}$, $u_s$ is a modest 80~$\mu$m/s and $u_n$ is nearly four orders
of magnitude smaller.

\subsection{Thermodynamic formalism}

The isothermal condition allows an effective thermodynamic description 
of the finite ${\bf Q}$ state~\cite{HM65}.  Although nonequilibrium scaling 
does not rely on this, a more intuitive description of a number of phenomena 
special to superfluidity results, and so it is worthwhile presenting 
the theory in this context.  

In the frame of reference moving with the normal fluid (in which there is 
no heat flow), the differential of the free energy density 
$F(T,{\bf U}_s)$ may be written,
\begin{equation}
dF = -SdT + {\bf J}_s \cdot d{\bf U}_s,
\label{2.4}
\end{equation}
in which ${\bf U}_s = {\bf u}_s - {\bf u}_n$ and 
\begin{equation}
{\bf J}_s = \left({\partial F \over \partial {\bf U}_s} \right)_T 
= \rho_s {\bf U}_s,
\label{2.5}
\end{equation}
with the second equality serving as a {\it definition} of $\rho_s(T,U_s)$
when $U_s$ is not small.  Near $T_\lambda$ the smallness of ${\bf u}_n$ 
implies that ${\bf U}_s \simeq {\bf u}_s$ and ${\bf J}_s \simeq {\bf j}_s$.  
In this frame one has the defining relation ${\bf U}_s = (\hbar/m) 
\nabla \phi$, where $m$ is the $^4$He atomic mass, and
\begin{equation}
\phi({\bf r}) = {m \over \hbar} {\bf U}_s \cdot {\bf r}
\label{2.6} 
\end{equation}
is the phase of the superfluid order parameter $\psi = |\psi| e^{i\phi}$.
Thus, $\psi$ rotates in the complex plane as one moves along the
direction ${\bf U}_s$, generating a kind of helical structure.  The 
effective thermodynamic description (\ref{2.4}) relies on the existence 
of a time-independent $\psi$ in the presence of a finite phase gradient.  
In fact, thermally nucleated phase slips in the helical structure 
(interpreted as tunneling between different metastable local minima 
of the free energy) lead to small temperature gradients and decay of 
superflow.  However, the decay time is extremely large at low heat 
currents and temperatures not too close to $T_\lambda$~\cite{LF67}.  
The thermodynamic description (\ref{2.4}) is valid on time scales 
smaller than this.  As ${\bf U}_s$ increases, both the order parameter 
magnitude $|\psi(U_s,T)|$ and the superfluid density are suppressed 
relative to their equilibrium values at $U_s = 0$, $Q = 0$.

Variations at constant ${\bf Q} = -T S {\bf J}_s$ are most conveniently 
performed by defining the Legendre transformed free energy $\Phi(T,{\bf J}_s) 
= F - {\bf J}_s \cdot {\bf U}_s$ with differential
\begin{equation}
d\Phi = - SdT - {\bf U}_s \cdot d{\bf J}_s.
\label{2.7}
\end{equation}
Close to $T_\lambda$, where $TS \simeq T_\lambda S_\lambda$ is essentially 
constant, variations at constant ${\bf Q}$ are asymptotically the same as 
those at constant ${\bf J}_s$.  For example, the specific heat at fixed 
${\bf Q}$ may be taken as
\begin{equation}
C_Q = T \left({\partial S \over \partial T}\right)_{{\bf J}_S},
\label{2.8}
\end{equation}
where $S(T,{\bf J}_S) = -(\partial \Phi/\partial T)_{{\bf J}_s}$.  With
the above observation in mind, we shall henceforth treat ${\bf J}_s$ and
${\bf Q}$ as differing only by a constant factor.

\subsection{Nonequilibrium scaling}

The fact that ${\bf U}_s$ and ${\bf J}_s$ (or ${\bf Q}$) may be treated 
as thermodynamically conjugate variables has important consequences for 
the structure of the thermodynamic functions near $T_\lambda$~\cite{F73}.  
We proceed by analogy with the conjugate variables $h$ and $m$ at conventional 
critical points, where $h$ is the external magnetic field and $m$ the 
magnetization at a Curie point, or $h$ is the difference from critical 
pressure and $m$ the difference from critical density at a liquid-vapor 
critical point.  There the free energy $A(T,h)$, analogous to $\Phi$, 
has differential $dA = - S dT - m dh$ and its singular part 
$A_s$ obeys an {\it asymptotic scaling form} [see, e.g., \cite{F83}],
\begin{equation}
A_s(T,h) = E_0 |\epsilon|^{2-\alpha} 
{\cal A} \left({D_0 h \over |\epsilon|^\Delta} \right),
\label{2.9}
\end{equation}
in which $\alpha$ is the specific heat exponent, $\Delta$ is the
``gap exponent,'' ${\cal A}(x)$ is a universal scaling function, and $E_0$, 
$D_0$ are nonuniversal scale factors, specified uniquely via, say, 
the normalizations ${\cal A}(0) = {\cal A}'(0) = 1$.  There are actually 
two scaling functions, ${\cal A}_\pm(x)$ for $\pm \epsilon > 0$, 
but we shall primarily be interested in the ordered phase $\epsilon < 0$ 
and consider only ${\cal A}_- \equiv {\cal A}$.

\begin{figure}[t]
\centerline{\psfig{figure=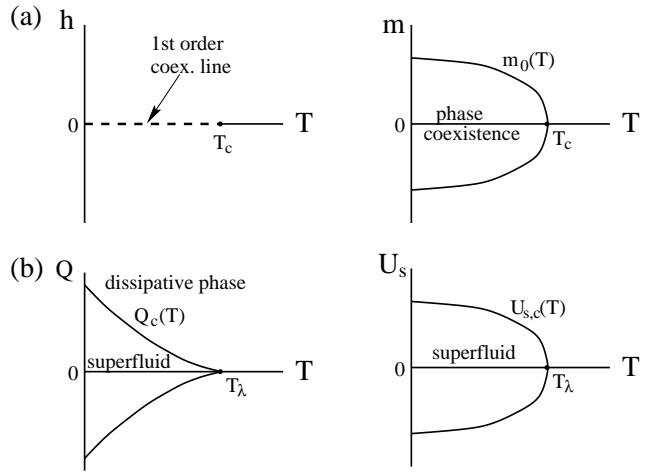,width=3.3in,angle=0.0}}
\vskip0.1in
\caption{Comparison of phase diagrams in (a) conventional $T$-$h$ and $T$-$m$ 
spaces and (b) superfluid $T$-$Q$ and $T$-$U_s$ spaces.  Here, $m_0(T)$ is the 
spontaneous magnetization, or the liquid density minus the critical density.  
The critical lines $Q_c(T)$ and $U_{s,c}(T)$, enclose the region of stable 
superflow, which corresponds also to the region of validity of the thermodynamic 
description, and like the thermodynamic description itself are sharp only in 
the absence of phase slips.  The different shapes of these curves near $T_\lambda$
are determined by the fact that $\Delta_Q > 1$, while $\beta_Q < 1$ [see equations
(\ref{2.18})--(\ref{2.21}) below].  The nature of the critical behavior as $Q$ 
approaches $Q_c$ from below, and the nature of the inhomogeneous dissipative phase 
for $Q > Q_c$, will be discussed in later sections.  A major difference between (a) 
and (b) is the lack of a first order line below $T_\lambda$ in the latter.  Thus, 
$h$ vanishes throughout the phase coexistence region in the $T$-$m$ plane, whereas 
$Q \propto J_s = \rho_s U_s$ varies continuously throughout the superfluid phase 
in the $T$-$U_s$ plane.}
\label{fig1}
\end{figure}
 
Similarly, the singular part of the superfluid free energy, $\Phi_s$, is 
expected to obey the scaling form,
\begin{equation}
\Phi_s(T,Q) = A_0 |\epsilon|^{2-\alpha} 
Y \left({Q \over Q_0 |\epsilon|^{\Delta_Q}} \right),
\label{2.10}
\end{equation} 
in which $\alpha \simeq -0.013$~\cite{lpe96} is again the usual equilibrium 
specific heat exponent, $\Delta_Q$ is the gap exponent for $Q$, $Y(y)$ is 
the $\epsilon < 0$ universal scaling function, and $A_0$, $Q_0$ are nonuniversal 
scale factors, specified uniquely via, say, the normalizations 
$Y(0) = Y^{\prime\prime}(0) = 1$ [$Y(y)$ must be an even function of $y$ 
due to the obvious symmetry under sign reversal of $Q$].  

The derivative of (\ref{2.9}) with respect to $h$ yields 
\begin{equation}
m(T,h) = - E_0 D_0 |\epsilon|^\beta {\cal A}'
\left({D_0 h \over |\epsilon|^\Delta} \right)
\label{2.11}
\end{equation}
where the prime denotes derivative with respect to argument and the order 
parameter exponent $\beta$ obeys the scaling relation $\beta = 2-\alpha-\Delta$.  
The second $h$ derivative yields the order parameter susceptibility
(compressibility, in the case of a liquid-vapor critical point)
\begin{equation}
\chi(T,h) = - E_0 D_0^2 |\epsilon|^{-\gamma} 
{\cal A}^{\prime\prime}\left({D_0 h \over |\epsilon|^\Delta} \right)
\label{2.12}
\end{equation}
with $\gamma = \alpha + 2\Delta - 2$, which then yields the famous
Essam-Fisher scaling law $\alpha + 2\beta + \gamma = 2$~\cite{F83}.

Similarly, the derivative of $\Phi_s$ with respect to $J_s$ yields 
$U_s$ in the form \cite{HD92b}
\begin{equation}
U_s(T,Q) = -B_0 |\epsilon|^{\beta_Q} 
Y'\left({Q \over Q_0 |\epsilon|^{\Delta_Q}} \right),
\label{2.13}
\end{equation}
where $B_0 = A_0 T_\lambda S_\lambda/Q_0$, and one has the generalized
order parameter exponent scaling relation
\begin{equation}
\beta_Q = 2-\alpha-\Delta_Q.
\label{2.14}
\end{equation}
The equilibrium superfluid density enters the free energy $F$ via a term 
$\Delta F_s = {1 \over 2} \rho_s U_s^2$ for small $U_s$.  The Legendre 
transform yields a term $\Delta \Phi_s = -J_s^2/2\rho_s$ for small $J_s$, 
and the inverse of the superfluid density now appears in the theory as a 
generalized susceptibility:
\begin{eqnarray}
{1 \over \rho_s(T,Q=0)} &=& -\left(\partial^2 \Phi_s 
\over \partial J_s^2 \right)_{T,J_s=0}
= \left(\partial U_s \over \partial J_s \right)_{T,J_s=0} 
\nonumber \\
&=& -{R_0 Y^{\prime\prime}(0) \over |\epsilon|^{\gamma_Q}}
\label{2.15}
\end{eqnarray} 
where $R_0 = A_0 (T_\lambda S_\lambda/Q_0)^2$ and the generalized
susceptibility exponent $\gamma_Q$ obeys the Essam-Fisher relation~\cite{F83},
\begin{eqnarray}
&&\gamma_Q = \zeta = 2\Delta_Q + \alpha -2
\label{2.16} \\
&&\alpha + 2\beta_Q + \gamma_Q = 2
\label{2.17}
\end{eqnarray}
Generally the gap exponent is independent of $\alpha$ and must be separately
determined.  However, in the superfluid problem the Josephson relation [see, 
e.g., \cite{PJF74} and references therein] yields $\zeta = 2 - \alpha - 2\nu 
= (d-2) \nu$.  Here $\xi \approx \xi_0/|\epsilon|^\nu$ with, in dimension $d=3$, 
$\nu = \zeta \simeq 0.671$ and $\xi_0 \simeq 3.4$\AA~\cite{GA92}, describes the 
divergence of the superfluid coherence length, $\xi = (m^2 k_B T/\hbar^2 
\rho_s)^{1/(d-2)}$, and the second equality follows from the hyperscaling 
relation $2-\alpha = d\nu$~\cite{F73,F83}.  We therefore identify
\begin{equation}
\Delta_Q = 2-\alpha-\nu = (d-1)\nu.
\label{2.18}
\end{equation}
This scaling law implies that $Q$ scales with the cross-sectional 
area $\xi^{d-1}$ of a correlation volume $\xi^d$:  $Q$ becomes significant
when the power incident on a correlation area is of order $Q_0 
\xi_0^{d-1}$.\footnote{One may estimate $Q_0$ as follows.  Two scale factor 
universality yields a form $C_s = k_B (R_\xi/\xi)^d/\alpha |\epsilon|^2$ for 
the singular part of the equilibrium specific heat below $T_\lambda$, where the 
hyperuniversal ratio $R_\xi \simeq 0.90$ in $d=3$ \protect\cite{HAHS76}.  This 
must match the scaling form (\protect\ref{2.10}) at $Q=0$ and determines
$A_0 |\epsilon|^{2-\alpha} = [k_B T_\lambda/\alpha (1-\alpha) (2-\alpha)] (R_\xi/\xi)^d$ 
[with the choice $Y(0) = 1$].  Equating the quadratic terms $A_0 |\epsilon|^{2-\alpha} 
(Q/Q_0 |\epsilon|^{\Delta_Q})^2 = -J_s^2/\rho_s$ [with the choice $Y^{\prime\prime}(0) 
= 1$], one obtains finally $Q_0 = T_\lambda S_\lambda \sqrt{-A_0 \rho_0}$, and
$Q_0 \xi_0^{d-1} = mk_B T_\lambda^2 S_\lambda R_\xi^{d/2}/\hbar \sqrt{-\alpha (1-\alpha)
(2-\alpha)}$ (using $\rho_0= m^2 k_B T_\lambda/\hbar^2 \xi_0^{d-2}$).  Substitution 
of experimental numbers yields $A_0 \simeq -21$J/cm$^3$ and $Q_0 \simeq 30$kW/cm$^2$.}  
From (\ref{2.14}) one obtains
\begin{equation}
\beta_Q = \nu,
\label{2.19}
\end{equation}
This relation has the interpretation that $(m/\hbar) {\bf U}_s$, which has 
dimensions of inverse length, scales with the inverse correlation length $\xi^{-1}$:
the phase gradient has a significant effect when its wavelength $2\pi\hbar/m 
U_s$ becomes comparable to $\xi$.  Note that, more typically, one begins with 
the latter assumption and reverses the above argument to {\it derive} the 
Josephson relation.

\begin{figure}
\centerline{\psfig{figure=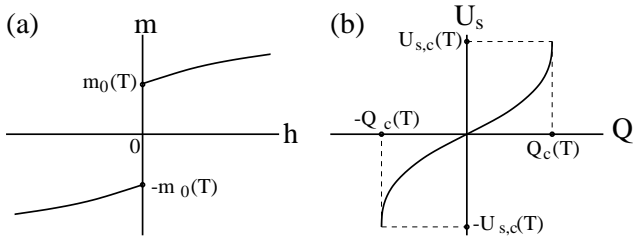,width=3.3in,angle=0.0}}
\vskip0.1in
\caption{Comparison of the isothermal equation of state (a) in the conventional 
$h$-$m$ plane and (b) in the superfluid $Q$-$U_s$ plane.  At $h=0$ there is a 
first order transition between up and down magnetized states, or between liquid 
and vapor states.  As $|h|$ increases, $|m(T,h)|$ increases as well.  As $|Q|$ 
increases $|U_s|$ increases until the superfluid breaks down at $|Q|=Q_c(T)$, 
at which point $U_s(Q,T)$ has a square root cusp.  At this same point the 
specific heat at constant $U_s$ also has a square root cusp~\protect\cite{HD94}, 
while the specific heat at constant $Q$ has an inverse square root 
divergence \protect\cite{cq96}.}
\label{fig2}
\end{figure}

Just as arbitrarily small $h$ smears the singular behavior near a Curie 
point, one expects an arbitrarily small $Q$ to either smear or drastically 
alter the lambda point critical behavior.  This is consistent with
$\Delta_Q > 0$, implying that the scaling argument $y$ diverges 
as $|\epsilon| \to 0$ at any finite $Q$, serving to define $Q$ as a 
{\it relevant perturbation} to the lambda point~\cite{F73}.  In Fig.~\ref{fig1} 
a schematic of the expected phase diagram is shown, contrasting it with 
that for a conventional critical point.  The lines $Q_c(T)$ and $U_{s,c}(T)$
are the boundaries beyond which superfluidity breaks down:  for $Q > Q_c$,
due to suppression of the order parameter and superfluid density, the heat 
current is too large for the superfluid to support isothermal heat transport, 
and a nonequilibrium phase transition to a new dissipative phase occurs.  
The nature of this phase will be discussed in Sec.\ III below.  In Fig.~\ref{fig2}
we sketch the isothermal equations of state for the conventional and superfluid
systems.  At a fixed subcritical temperature the conventional system displays the
usual first order jump in $m$ at $h=0$ between up and down magnetized states,
or between liquid and vapor states.  The superfluid system displays a continuous
variation of $U_s$ with $Q$, but at $|Q|=Q_c(T)$ encounters the boundary between
superfluid and dissipative phases.  At this point $U_s(Q)$ has a square root 
cusp~\cite{HD94}, corresponding to a strong suppression of $\rho_s$ and a divergent 
susceptibility $(\partial U_s/\partial J_s)_T$.  This singularity is exhibited 
in the scaling form (\ref{2.10}) as square root cusps in $Y(y)$ at some finite 
value $y=\pm y_c$.  This yields the predictions 
\begin{eqnarray}
Q_c(T) &\approx& Q_0 y_c |\epsilon|^{\Delta_Q}
\label{2.20} \\
U_{s,c}(T) &\approx& B_0 Y'(y_c) |\epsilon|^{\beta_Q} 
\propto Q_c(T)^{1/\delta_Q},
\label{2.21}
\end{eqnarray} 
in which 
\begin{equation}
\delta_Q = \Delta_Q/\beta_Q = 1 + \gamma_Q/\beta_Q.
\label{2.22}
\end{equation}
The latter relation generalizes the Widom scaling law
$m(T_c,h) \propto h^{1/\delta}$ with $\delta = \Delta/\beta = 1+\gamma/\beta$
which follows from (\ref{2.11}) in the limit $\epsilon \to 0$ [the asymptotic 
behavior ${\cal A}'(x) \sim x^{\beta/\Delta}$ as $x \to \infty$ is required for 
consistency]~\cite{F83}.  From (\ref{2.16}) and (\ref{2.19}) one obtains
explicitly $\delta_Q = d-1$.  As will be discussed in Sec.\ V, the 
specific heats at constant $U_s$ and $Q$ are also singular at $U_{s,c}(T)$ and
$Q_c(T)$~\cite{HD94,cq96,HD96}.

The coherence length itself obeys a scaling form
\begin{equation}
\xi(T,Q) = \xi_0 |\epsilon|^{-\nu} 
\Xi\left({Q \over Q_0 |\epsilon|^{\Delta_Q}} \right),
\label{2.23}
\end{equation}
with $\Xi(0) = 1$.  Considering variations of $\xi$ at a fixed value of 
$y = Q/Q_0 |\epsilon|^{\Delta_Q}$, one may write
\begin{eqnarray}
\xi &=& \xi_0 (Q/Q_0)^{-\nu_Q} y^{\nu_Q} \Xi(y) 
\label{2.24} \\
\nu_Q &\equiv& \nu/\Delta_Q = 1/(d-1),
\label{2.25}
\end{eqnarray}
serving to define an exponent describing the characteristic variation
of $\xi$ with ${\bf Q}$.  Attempts to verify the scaling 
relations (\ref{2.20}) and (\ref{2.25}), with $\Delta_Q = 2\nu$ and $\nu_Q 
= 1/2$ in $d=3$, will be discussed in later sections.

In Table \ref{table1} we summarize the various critical exponents we have 
defined, along with their values in general dimension $d$ and in $d=3$.

\begin{table}
\centerline{\(\begin{array}{ccr}
{\rm exponent} & {\rm general}\ d & d=3 \\ \hline
\nu & 1/2 \leq \nu(d) \leq \infty & 0.671 \\ 
\alpha & 2-d\nu & -0.013 \\ 
\zeta & 2-\alpha-2\nu = (d-2) \nu & 0.671 \\ 
\Delta_Q & 2-\alpha - \nu = (d-1)\nu  & 1.342 \\
\beta_Q & \nu & 0.671 \\ 
\gamma_Q & \zeta & 0.671 \\ 
\delta_Q & 1+\gamma_Q/\beta_Q = d-1 & 2 \\
\nu_Q & \nu/\Delta_Q = 1/(d-1) & 1/2 \\
\end{array}\)}
\vskip0.1in
\caption{Equilbrium and nonequilibrium critical exponents and their values.  
Note that no exact expression for the correlation length exponent $\nu(d)$ is 
known, aside from the boundary values $\nu(2) = \infty$ and $\nu(d \geq 4) = 
{1 \over 2}$, but all other exponents are either given exactly or in terms of 
$\nu$.  There are other independent exponents (e.g., the critical correlation 
exponent $\eta \simeq 0.02$), but they happen not to be involved in any of 
the experiments we discuss.  Expressions in the center column involving $d$ 
explicitly require hyperscaling, and are therefore valid only for $d \leq 4$.  
For $d \geq 4$ the exponents all stick at their mean field values, and may
be determined by substituting $\nu = {1 \over 2}$, $\alpha=0$ into the 
non-explicitly $d$-dependent forms of the scaling relations.}
\label{table1}
\end{table}

As a final comment, we note that verifications of certain dynamical scaling 
laws in $^4$He are complicated by a fundamental problem associated with the 
existence of very slowly convergent, nonasymptotic dynamic effects.  Scaling 
forms like (\ref{2.10}) implicitly assume that $|\epsilon|$ is sufficiently 
small that all {\it irrelevant} scaling variables may be ignored.  It turns 
out that a certain \emph{Wegner exponent} $\omega$ controls the relevance of
{\it mass and heat diffusion} on the dynamic critical behavior \cite{HH77} 
through an additional scaling variable $w \equiv \gamma_0 |\epsilon|^\omega$, 
where $\gamma_0$ is a parameter of order unity in the Model F equations 
\cite{HH77} (describing the near-critical dynamics of $^4$He) which couples
entropy and density fluctuations to order parameter fluctuations.\footnote{The 
Hohenberg-Halperin Models A--J classify the different dynamical universality 
classes encountered most commonly in physical systems.  The dynamical behavior 
depends qualitatively, for example, on whether or not the order parameter is a 
conserved density, and whether or not it is coupled to other conserved densities 
(like mass or energy).  The Model F equations are the simplest possible 
description of a non-conserved two-component order parameter (the real and 
imaginary parts of $\psi$) coupled to heat and mass flow, and are believed 
to provide an asymptotically exact description of the universal critical dynamics 
of superfluid $^4$He.}  The variable $w$ does not appear in any equilibrium 
scaling function (where such fluctuations may be completely ``integrated out'' 
of the partition function), but does appear in those involving dissipative 
transport, e.g., that of the thermal conductivity \cite{HD92}.  Since 
$\omega \simeq 0.008$ \cite{D91} is very small at the superfluid transition 
this variable vanishes extremely slowly as $|\epsilon| \to 0$:  $\epsilon = 
10^{-125}$ leads only to $|\epsilon|^\omega = 0.1$.  The scaling function 
then has a very slow parametric dependence on $w$, leading to ``quasi-scaling'' 
\cite{HD92} of the heat conductivity, and hence to an apparent slow variation 
of the associated critical exponent with $w$.  Measured dynamic critical 
exponents affected in this way may typically be expected to differ from their 
true asymptotic values by 10-20\% \cite{DAS88}.

The scaling phenomena considered here, however, are mainly those
associated with properties of isothermal superfluids, which though
out of equilibrium, are nevertheless argued to behave as equilibrium 
thermodynamic systems.  To the extent that this is true (i.e., to the 
extent that vortex excitations can be ignored), the Model F equations 
again produce a thermodynamic-type partition function with entropy/density 
fluctuations integrated out, and finite ${\bf Q}$ enforced by an imposed 
uniform helical twist in the order parameter \cite{HD94}.  The scaling 
variable $w$ will again not appear, and one expects that quasi-scaling will 
be absent in (\ref{2.10}) and from all quantities derived from it, and hence 
that the exponents in Table~\ref{table1} will exhibit experimentally their 
predicted values.  On the other hand, the phase boundaries
in Fig.~\ref{fig1}(b) are defined by the \emph{onset} of dissipative 
transport, with divergent fluctuations in the local heat current as they 
are approached.  Such fluctuations lead to vortex creation, 
decay of superflow, breakdown of the effective equilibrium description, 
and the reappearance of the variable $w$.  In some sense $w$ must control 
the ``fuzziness'' of the boundary $Q_c(T)$, and 
an experimental test of the relations (\ref{2.20}) may exhibit
quasi-scaling even if, for $y$ sufficiently less than $y_c$, the scaling 
function $Y(y)$ does not.  This issue is potentially testable by
the specific heat measurements discussed in Sec.\ V.  Unfortunately,
as exhibited in Fig.~\ref{fig10} below, present data are constrained
to lie sufficiently far below $Q_c(T)$ that both asymptotic and nonasymptotic
forms fit equally well.

\section{The nonequilibrium superfluid-normal interface} 

\subsection{Interface statics}

\begin{figure}
\centerline{\psfig{figure=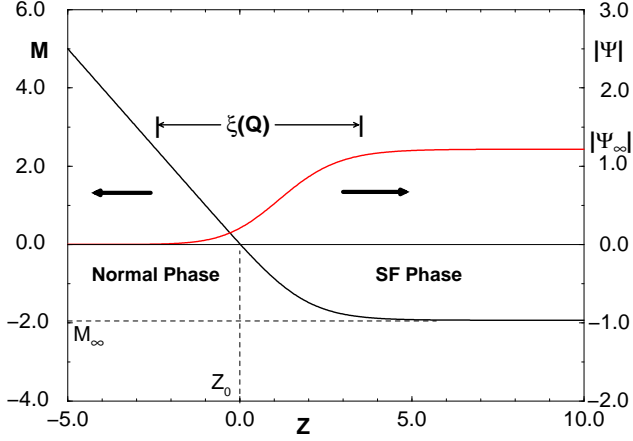,width=3.0in,angle=-90}}
\vskip0.1in
\caption{Scaled temperature and order parameter profiles $M(Z)$ and
$\Psi(Z)$, computed within the mean field approximation \protect\cite{O83,WPMM98}.
Here $Z=z/l(Q)$, $|\Psi| \propto |\psi| l(Q)$ and $M \propto (T-T_0)l^2(Q)$, where
$T_0$ is the mean field transition temperature and $l(Q) \propto Q^{-1/3}$ is
essentially the correlation length (\protect\ref{2.24}) in the mean field 
approximation.  With these scalings the profiles are independent of $Q$.}
\label{fig3}
\end{figure}

Consider a cylindrical cell with a heat current ${\bf Q}$ driven along its
axis, labeled by coordinate $z$, in which the up-stream endwall, $z=0$, 
has $T(0) > T_\lambda$.  In the normal phase heat is transported by thermal 
conduction, which at sufficiently small $Q$ is described by the Fourier law
\begin{equation}
Q = -\kappa \partial_z T,
\label{3.1}
\end{equation}
where $\kappa(T) \approx \kappa_0 \epsilon^{-\mu}$ is the thermal conductivity,
predicted to diverge at $T_\lambda$, consistent with its infinite value at all
$T < T_\lambda$.  Experimentally one finds \cite{DZM86,TA85,soc97} $\kappa_0
\simeq 12\mu$W/cm$^2$ and $\mu \simeq 0.44$.  In a very tall cell in which
$T$ varies substantially along its length, one may view (\ref{3.1}) as locally
valid with $\kappa(z) = \kappa[T(z)]$ so long as $T(z)$ remains sufficiently
far above $T_\lambda$.  ``Sufficiently far above $T_\lambda$'' may be quantified
by the condition that the temperature drop across a coherence length be much
smaller than the deviation from $T_\lambda$:
\begin{equation}
\xi |\nabla T| \ll T-T_\lambda ~~\Rightarrow~~ 
{Q \xi \over \kappa \epsilon T_\lambda} \ll 1.
\label{3.2}
\end{equation}
Putting in $^4$He parameters, this requires
\begin{equation}
\epsilon \gg 6 \times 10^{-8} 
\left({Q \over 1\mu{\rm W}/{\rm cm}^2} \right)^r,
~~r \equiv {1 \over 1 + \nu - \mu} \simeq 0.81
\label{3.3}
\end{equation}
As $z$ increases, $T(z)$ will eventually violate (\ref{3.3}), and one enters 
a region of {\it nonlinear heat transport}.  In effect, $\kappa$ becomes a
strong function of $Q$ in this region.  Moreover, as illustrated in Fig.~\ref{fig3}, 
at some position $z_0$, $T(z_0) = T_\lambda$, and the system enters the 
superfluid phase for $z > z_0$:  a nonequilibrium superfluid--normal fluid 
interface is generated.  Far downstream from this interface, $T(z)$ levels 
out at a temperature $T_\infty(Q)$, one of the nonequilibrium thermodynamic 
states discussed in Sec.\ II above.  Correspondingly, the order parameter 
magnitude $|\psi(z)|$, which effectively vanishes for $z < z_0$, grows in 
the interface region and saturates at a value $|\psi_\infty(Q)|$ for 
$z \gg z_0$.\footnote{In fact $T(z)$ continues to decrease slowly with 
$z$ in the superfluid phase due to rare phase slip events \cite{H99a,H99b}, 
so $T_\infty(Q)$ and $|\psi_\infty(Q)|$ are not sharply defined, but we 
shall not dwell on this complication here.}  

The interface is the region over which the mode of heat transport 
converts from conduction to superfluid counterflow.  Its characteristic width 
$\xi(Q)$, through which $T$ and $|\psi|$ vary substantially (Fig.~\ref{fig3}),
is expected to scale according to (\ref{2.24}).  We may estimate this width
semi-quantitatively by defining a reduced temperature scale $\epsilon_Q$
through equality in (\ref{3.2}) and (\ref{3.3}), and defining 
\begin{equation} 
\xi(Q) \equiv \xi(\epsilon_Q) \approx 24 
\left({Q \over 1~\mu{\rm W}/{\rm cm}^2} \right)^{-r \nu}~\mu{\rm m},
~~r\nu \simeq 0.54.
\label{3.4}
\end{equation}
The scaling relation (\ref{2.25}) predicts $r\nu = \nu_Q$ ($=1/2$ in $d=3$), 
requiring the scaling relation $\mu = 1-\zeta$ ($\simeq 1/3$ in $d=3$) 
\cite{HH77}.  The difference between experimental and theoretical values is
presumably due to the quasi-scaling effect described at the end of Sec.\ II.
Quantitative renormalization group based predictions for the full 
temperature profile $T(z,Q)$, explicitly including the quasi-scaling phenomenon 
\cite{HD91,HD92}, are limited to the normal fluid region $z < z_0$ where the vanishing 
of $\psi$ greatly simplifies the calculations.  Less quantitative predictions for 
the entire profile are limited to mean field \cite{O83} (which ignores critical
fluctuations) or large-$N$ \cite{H99a,H99b} (which replaces the single complex 
order parameter with an $N$-component vector) approximations.

An experimental measurement of the temperature profile $T(z,Q)$ within the 
interface region for a sequence of different $Q$ would allow a detailed 
exploration of near-critical, nonlinear heat transport.  It would also provide 
an experimental test of the scaling prediction (\ref{2.25}).  Such a measurement 
requires a regime in which $\xi(Q)$ is larger than the width $W$ of the thermometer.  
Present technology places a limit $W \agt 50~\mu$m.  From (\ref{3.4}), heat currents 
below $1~\mu$W/cm$^2$ are required.  Present technology allows controlled heat 
currents down to about 1~nW/cm$^2$, leading to $\xi \approx 1$~mm, so at first 
sight such an experiment appears feasible.  

In fact, Earth's gravity $g_e$ places a fundamental limit on the maximum possible 
interface width.  As described in Sec.\ I, gravity produces an equilibrium ($Q=0$) 
superfluid-normal fluid interface with width of order $\xi(g_e) = 100~\mu$m.  From 
(\ref{3.4}), $\xi(Q) = \xi(g_e)$ for $Q \simeq Q_g \equiv 70$~nW/cm$^2$.  For $Q$ 
of order $Q_g$, gravity will begin to have a strong effect on the nonequilibrium 
interface \cite{unm98}, and for $Q \ll Q_g$ the heat current will be a small
perturbation on the equilibrium interface.  Thus $\xi(Q)$ will saturate at 
$\xi(g_e)$ as $Q \to 0$ and the regime $\xi(Q) \gg W$ is unattainable on Earth.

For this reason the critical dynamics experiment (DYNAMX) is currently being
prepared for the microgravity environment of the International Space Station, 
with flight planned for 2004.  Heat currents as low as 5~nW/cm$^2$ will be 
used.  The temperature profile $T(z,Q)$ will be measured to subnanoKelvin
resolution by slowly moving the interface past a fixed thermometer.  The 
latter is accomplished by removing heat from the downstream end of the cell
slightly more slowly than it enters the upstream end, with the effect that
the normal phase slowly invades the cell, thus translating the interface.

\subsection{Transition from thermodynamic to interface state}

We have discussed three possible classes of steady nonequilibrium states:
(a) conducting normal states with static temperature profile determined 
by (\ref{3.1}); (b) isothermal thermodynamic states with steady superfluid
counterflow determined by (\ref{2.1}); and (c) states with a nonequilibrium 
interface forming a ``conversion boundary'' between states of type (a) and (b). 
Transitions between (a) and (c) occur continuously:  if
state (a) is cooled to the point where the downstream endwall temperature
passes through $T_\lambda$, an interface will form out of that endwall and
steadily move upstream.  Conversely, as heat is added to state (c), the 
interface will move downstream until the interface disappears into the 
endwall.

Similarly, the transition from (c) to (b) is continuous.  The interface
will disappear into the upstream endwall as heat is extracted, yielding
an isothermal thermodynamic state at temperature $T_\infty(Q)$.  Further
extraction of heat will cause the temperature to drop below $T_\infty(Q)$.

\begin{figure}
\centerline{\psfig{figure=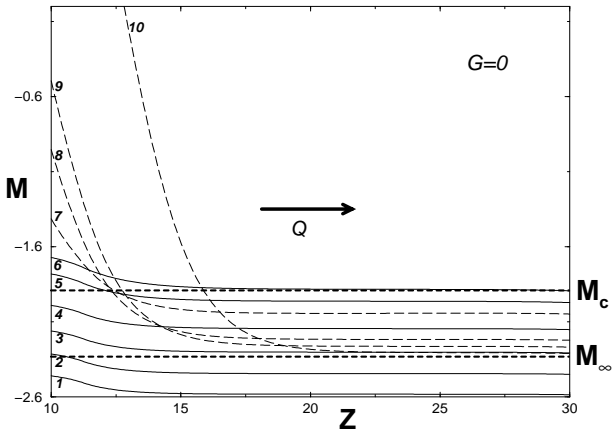,width=3.2in,angle=270}}
\vspace{0.1in}
\caption{For bulk scaled temperature $M > M_c$ (see caption to 
Fig.~\ref{fig3}) the uniform superfluid state becomes unstable to a state 
with an interface \protect\cite{WM00}.  Curves 1-6 (solid) show consecutive 
{\it stable} (time independent) superfluid temperature profiles obtained 
in the mean field approximation upon slowly varying the right-hand wall 
temperature.  Curves 7-10 show snapshots of the time evolution, computed 
within a simple one-dimensional model, initiated by a series of phase slips 
(not shown) after the right-hand wall temperature is raised slightly above 
that of curve 6.  The key feature is the net temperature drop (from $M_c$ 
to $M_\infty$) observed in the bulk superfluid.  Figure reprinted from 
\protect\cite{WM00} courtesy of the Journal of Low Temperature Physics.}
\label{fig4}
\end{figure}

The nature of the transition from (b) to (c) is less clear.  At issue is
whether the bulk superfluid, in the absence of an interface, recognizes
$T_\infty(Q)$ as a special temperature.  Within the mean field approximation,
the answer is no \cite{O83}:  the thermodynamic state is stable up to a 
temperature $T_c(Q)$, with $T_\lambda > T_c(Q) > T_\infty(Q)$, whose 
functional inverse $Q_c(T)$ we might identify with the boundary 
in Fig.~\ref{fig1}(b), at which the superfluid density is suppressed to the 
point where it is incapable of supporting the heat current.  As shown in 
Fig.~\ref{fig4}, when the thermodynamic state is heated above $T_c(Q)$ a 
complicated dynamics results, with the system finally settling down into 
a state with an interface.  Since $T_\infty(Q) < T_c(Q)$, the superfluid 
side actually \emph{cools}, with a compensatory heating of the normal side.  
The final position of the interface is determined from energy conservation.

The question of whether or not a first order transition 
from (b) to (c) survives in some form beyond the mean field approximation is 
subtle, and is probably a question of time scale.  The same thermal nucleation 
of vortices that leads to a small temperature gradient in the superfluid phase
presumably also leads to a continuous (b) to (c) transition, with the interface
entering continuously from the upstream endwall, if the experimental heating
rate is infinitesimally slow.  Any finite heating rate may, however, allow
a ``superheating'' of (b), inducing an apparent first order transition to (c)
nucleated by a random vortex creation event.  Such an effect would be analogous
to superheating and supercooling effects at conventional first order transitions,
the boundary $Q_c(T)$ being analogous to a spinodal line (defined roughly as the 
temperature at which a local free energy barrier separating the superheated 
or supercooled metastable state from the true equlibrium state disappears, 
allowing rapid conversion to the thermodynamically stable phase).

Experimental investigation of the (b) to (c) transition is complicated by 
boundary effects.  The singular Kapitza resistance leads to an additional
heating of the upstream endwall, which then acts as a nucleation point for vortices 
even when the temperature of the bulk superfluid is still below $T_\infty(Q)$ 
\cite{cq99}.  Clever cell designs that reduce the heat current near 
the endwall below that in the bulk may eventually allow experimental 
observation of a superheating effect.
 
\subsection{Interface dynamics}

Despite the greatly reduced gravity, the Space Station environment is less
than ideal for other reasons.  Vibrational noise (``g-gitter'') exists at a level 
of about $10^{-3} g_e$, and one might worry that such noise might couple strongly
to the interface, and perhaps destabilize it.  Understanding the effects
of acceleration noise requires an understanding of (a) the \emph{free dynamics} 
of the interface when it is perturbed away from its steady state, e.g., whether
or not the interface is even dynamically stable, or whether small perturbations
might undergo some kind of ``dendritic growth'', and (b) the manner in which 
vibration, or other perturbations, couple to, and perhaps amplify this motion 
\cite{WPMM98}.

Suppose that a slow variation, $z_0 = z_0({\bf r})$, with the transverse 
coordinates ${\bf r} = (x,y)$, is applied to the interface position and 
is subsequently released.  The problem is to derive an equation of motion
for $z_0({\bf r},t)$.  In order to motivate the interesting physical 
questions, it is useful to consider first the analogous problem of the 
motion of an \emph{equilibrium} interface between up and down domains 
of an Ising ferromagnet, with diffusive ``Model A'' dynamics \cite{HH77}.  
\emph{Surface tension} acts as a restoring force against perturbations away 
from a flat interface, and one may derive a \emph{diffusion equation} for 
the relaxation of long wavelength perturbations of the interface position:
\begin{equation}
(\partial_t - D \nabla^2) z_0({\bf r},t) = \eta({\bf r},t),
\label{3.5}
\end{equation}
in which the value of $D$ depends on the microscopic parameters in the model,
and the driving term $\eta$, vanishing for free relaxation, has been included 
for completeness.  {\it Thermal fluctuations} in the spins lead to a Gaussian
white noise form for $\eta$ with correlator $\langle \eta({\bf r},t) 
\eta({\bf r}',t') \rangle = \zeta_0 \delta({\bf r}-{\bf r}') \delta(t-t')$
where $\zeta_0$ depends on microscopic parameters and on temperature.  Using
this form one may compute the equal time variance in the interface position
\begin{eqnarray}
C({\bf r}-{\bf r}') &\equiv& \langle [z_0({\bf r},t)-z_0({\bf r}',t)]^2 \rangle
\nonumber \\
&=& {\zeta_0 \over 4\pi D} \ln \left({|{\bf r}-{\bf r}'| \over a_0} \right),
\label{3.6}
\end{eqnarray}
where $a_0$ is an atomic length scale.  One sees that fluctuations in the
interface position \emph{diverge} logarithmically with separation:  a famous 
result known as \emph{interface roughness}, encountered most frequently in
discussions of crystal facet shapes.  The interface is locally stable and 
flat, but globally wanders arbitrarily large distances.

The result (\ref{3.6}) shows that there are subtle physical issues, lying 
beyond the much simpler question of stability, arising because the interface 
breaks the continuous translation invariance of the system.  Whenever a 
continuous symmetry is broken, a Goldstone mode (namely, a slow dynamical
mode arising when the local value of the broken symmetry varies slowly in 
space) is generated, corresponding in this case to very slow relaxation of long 
wavelength perturbations of the interface position, $z_0 \propto e^{i{\bf k} 
\cdot {\bf r} - \lambda({\bf k}) t}$, with (\ref{3.5}) yielding $\lambda = Dk^2$.  
Such long wavelength ``modes'' are highly susceptible to thermal or externally 
generated noise spectrum $\zeta(k)$, and it is the convergence at small $k$ 
of the integral
\begin{equation}
\langle z_0({\bf r},t)^2 \rangle = \int {d^2k \over (2\pi)^2} 
{\zeta({\bf k}) \over {\rm Re}\lambda({\bf k})},
\label{3.7}
\end{equation}
that determines whether or not the interface is rough.  For the case 
$\zeta(k) \equiv \zeta_0$, (\ref{3.7}) diverges logarithmically, which 
same divergence is reflected in the correlator (\ref{3.6}).

The physics of the nonequilibrium superfluid-normal interface is very
different from that of the equilibrium magnetic interface.  As described,
the dynamics of the latter is purely diffusive, with the two bulk phases
on either side of it containing no slow dynamical modes of their own
(since the order parameter is non-conserved and is not coupled to any 
other conserved field).  In contrast, transport on the normal side of the 
$^4$He interface is controlled by slow heat diffusion, and transport on the 
superfluid side, controlled by superfluid counterflow, is essentially ballistic, 
and the dynamics of the interface itself therefore involves a very intricate 
coupling of \emph{three} different slow dynamical modes:  normal phase 
diffusion, superfluid phase counterflow, and the broken translational 
symmetry Goldstone mode of the interface itself.  To leading order one 
finds an equation of motion \cite{WPMM98}
\begin{equation}
(\partial_t^2 - c^2 \nabla^2) z_0({\bf r},t) = \eta({\bf r},t), 
\label{3.8}
\end{equation}
so that the interface supports \emph{traveling wave} excitations, with a 
well defined speed $c(Q)$ of the same order as the bulk second sound speed
(the speed with which perturbations in the order parameter travel on 
the superfluid side of the interface).
At next-to-leading order one finds that these excitations are \emph{singularly
damped}, with $\lambda = ick + Dk^{3/2}$, so that Re$\lambda \propto k^{3/2}$
rather than $k^2$ (as for bulk second sound waves) at small $k$.  Positivity
of Re$D$ establishes intrinsic dynamical stability of the interface.  One 
may interpret the enhanced damping as arising from the waves on the interface
``rubbing'' up against the normal phase.  Moreover, one finds that thermal
fluctuations enter via a spectrum $\zeta(k) \propto k^{5/2}$ for $\eta$,
vanishing strongly as $k \to 0$.  The physics of this result is related to
the fact that interface motion is a \emph{cooperative} phenomenon, involving
evanescent dynamics of the superfluid order parameter to a depth $\sim k^{-3/2}$ 
scaling as the $3/2$ power of the wavelength.\footnote{The evanscent depth on
the normal side is much smaller, scaling as $k^{-1/2}$.  For comparison, 
gravity waves on fluid surfaces yield fluid motion to a depth $\sim k^{-1}$ 
proportional to the wavelength.}  The microscopic thermal noise, which is white, 
must then be averaged over a similar volume to obtain its net effect on the mode,
leading eventually to the greatly reduced $\zeta(k)$ above.  In contrast, the
Ising interface moves by local spin flips and the microscopic noise is averaged
only over a microscopic region of width $a_0$ and $\zeta$ remains white.

The net result of the analysis is that the integral (\ref{3.7}) is strongly
convergent at $k=0$, and the superfluid-normal interface is globally flat.
This is good news for the Critical Dynamics Experiment (DYNAMX), where a 
rough interface would have led to substantial smearing of the temperature 
profile on a scale varying as the logarithm of the cell cross-section.

Vibrational acceleration noise couples to the interface in the same way
that Earth's gravity does, through the variation in the local $T_\lambda$ 
with pressure.  A slight change in $T_\lambda$ will cause the interface to 
translate (for acceleration normal to the interface) or tilt (for acceleration 
parallel to the interface) slightly, but if the change is \emph{oscillatory} 
it could resonate with one of the interfacial second sound modes, leading to 
a rapid growth in the interface motion.  A detailed examination of the expected 
frequency spectrum of the Space Station g-gitter, together with the discrete 
spectrum of standing wave modes allowed in the experimental cell within the 
planned temperature and heat current range, shows that such resonances may 
indeed occur, but that the singular damping is sufficiently strong that the 
interface oscillation amplitude should saturate at acceptably low levels 
\cite{DXDF27}.

The existence of the interfacial second sound mode has yet to be tested
experimentally.  This might be accomplished by applying a sequence of heat 
pulses to the cell sidewall near the interface and detecting a response
at the opposite sidewall.  Scattering of bulk second sound pulses off
the interface, with detection of the reflected pulses, might also provide
interesting information about the coupling of bulk and interface modes.

\section{The self-organized critical state} 

We have so far discussed phenomena in which optimal conditions occur in
the absence of gravity.  It transpires that there is a very interesting 
phenomenon in which gravity and heat current \emph{combine} to produce a 
new type of dynamical state.  The so-called \emph{self-organized critical} 
(SOC) state occurs in a cell which is heated from above, so that ${\bf g}$ 
and ${\bf Q}$ are parallel.\footnote{When $Q$ and $g$ are antiparallel
(heat from below) they cooperate to simply produce a sharper interface.}

Gravity depresses the lambda point $T_\lambda(z)$ with increasing depth
according to (\ref{1.1}), while heat current leads to decreasing temperature 
$T(z)$ with depth according to (\ref{3.1}).  The reduced temperature 
$\epsilon(z) \equiv [T(z) - T_\lambda(z)]/T_{\lambda,0}$, where 
$T_{\lambda,0}$ is, say, the bulk transition temperature if gravity 
were absent (and hence approximately the transition temperature at the 
top of the cell), contains a \emph{competition} between these two effects.
One might, in fact, imagine tuning $Q$ in such a way that $\epsilon(z)$ 
is independent of $z$ \cite{O87}:  the sample would apparently exist in an 
essentially homogeneous near-critical state.  In fact, it was argued 
\cite{MCH93,AL96} that the system actually ``self-organizes'' $T(z)$
in order to enforce a uniform $\epsilon$.  Assuming the validity of the 
Fourier law (\ref{3.1}), $\epsilon$ must be uniquely defined by
\begin{equation}
\kappa(\epsilon_{\rm SOC}) = Q/|\partial_z T_\lambda|,
\label{4.1}
\end{equation} 
where $\epsilon_{\rm SOC}$, a function of the ratio $Q/g$, is the reduced
temperature of the new state.  This state has recently been observed 
experimentally \cite{soc97}.

As an aside, we comment that the name SOC is motivated by similar 
self-tuning to a macroscopically homogeneous state under nonequilibrium 
conditions observed, for example, in ``sandpile'' models \protect\cite{BTW88}, 
and in fluid turbulence.  There, however, the self-organized state displays 
``avalanches'' (vortical eddies in the fluid), with a power law distribution 
of sizes, analogous to similar power law-distributed fluctuations observed at 
equilibrium critical points, but without the requirement that an external
parameter like temperature be tuned to obtain the critical state.  Although 
the ``SO'' part of SOC is justified for the $^4$He state, the ``C'' part is 
not since analogous critical power laws have yet to be demonstrated either 
theoretically or experimentally, as should become clear below.  The name, 
however, has stuck and we will not attempt to alter convention here.

The fact that $\kappa$ increases with decreasing $T$ ensures 
stability of the SOC state to small perturbations \cite{MCH93,AL96}.  
More specifically, an analysis of the heat diffusion equation in the normal 
phase \cite{WM00} shows that a perturbation $\epsilon({\bf x}) = 
\epsilon_{\rm SOC} + \delta \epsilon({\bf x})$ obeys an equation of motion 
whose solutions are decaying plane waves of the form
\begin{equation}
\delta \epsilon({\bf x},t) = \delta \epsilon({\bf q}) e^{-D_{\rm SOC} q^2 t}
e^{i{\bf q} \cdot ({\bf x} + c_{\rm SOC} {\bf \hat z} t)},
\label{4.2}
\end{equation}
with $D_{\rm SOC} = \kappa(\epsilon_{\rm SOC})/C_p(\epsilon_{\rm SOC})$ 
and $c_{\rm SOC} = -|\partial_z T_\lambda| \kappa'(\epsilon_{\rm SOC})/
T_{\lambda,0} C_p(\epsilon_{\rm SOC})$, where $C_p$ is the equilibrium 
specific heat at constant pressure.  Thus, in addition to the decay
controlled by the diffusion constant $D_{\rm SOC}$, there is an unexpected 
anisotropic propagation effect where the perturbation moves upstream at
speed $c_{\rm SOC}$.  For the reasonable value $Q = 50$~nW/cm$^2$ (see below)
one finds $c_{\rm SOC} \simeq 2.8$~mm/s, and the propagation effect should 
be experimentally observable for reasonable cell geometries \cite{WM00}.

\begin{figure}
\centerline{\psfig{figure=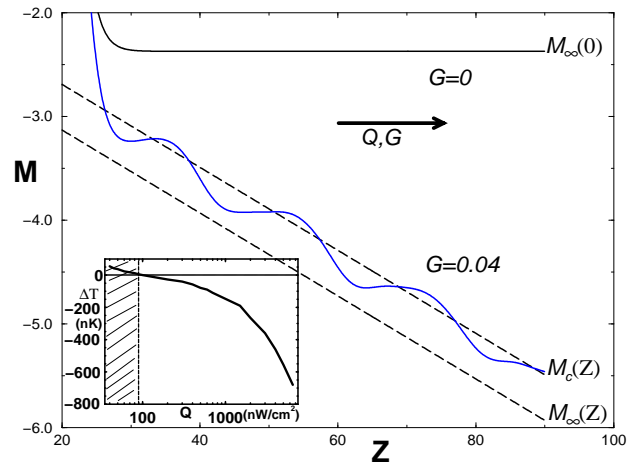,width=3.4in,angle=-90}}
\caption{Simulation of the SOC state using the same simplified 1-d model 
as in Fig.~\ref{fig4}.  For $G \propto g/Q = 0$ the temperature gradient 
in the normal phase ($Z \protect\alt 20$) gives way to an asymptotically 
isothermal superfluid phase ($Z \protect\agt 30$) with temperature 
$T_\infty(Q) < T_{\lambda,0}$.  For $G > 0$ the superfluid phase 
develops phase slips (vortices in 3-d), and a corresponding dynamic 
staircase structure in $T(z)$, roughly bounded between $T_\infty(z)$ 
and $T_c(z)$, to produce the SOC state.  The density of phase slips 
increases with $G$.  Inset:  experimental data replotted from Fig.~4 
of \protect\cite{soc97} showing the self-organization temperature, 
$\Delta T(Q) = T(Q,z) - T_{\lambda}(z)$.  Only for $Q \protect\alt 
100$nW/cm$^2$ (small shaded region) is the SOC state in the normal 
phase.  Figure reprinted from \protect\cite{WM00} courtesy of the 
Journal of Low Temperature Physics.}
\label{fig5}
\end{figure}

Since $\kappa$ diverges as $\epsilon \to 0$, (\ref{4.1}) implies 
that $\epsilon_{\rm SOC} \to 0$ as $Q \to \infty$.  However, 
for large $Q$ the Fourier law breaks down.  In particular 
$\kappa(T_\lambda,Q)$ is finite for $Q > 0$ \cite{HD92}, and the SOC 
state must therefore lie below $T_\lambda(z)$ for sufficiently 
large $Q > Q_{SOC}$.  This is consistent with experimental data 
\cite{soc97}, reproduced in the inset to Fig.~\ref{fig5}, which shows that 
$\epsilon_{\rm SOC} < 0$ for $Q \agt 100$~nW/cm$^2$.  Using (\ref{4.1}), 
along with (\ref{3.2}) defining the validity of the Fourier law, 
one may show that the previous theory is valid for \cite{O96,WM00}
\begin{equation}
6.1 \left(Q \over 100{\rm nW/cm^2}\right)^{(1+\nu)/\mu} 
\left(g \over g_e \right)^{1-(1+\nu)/\mu} \ll 1.
\label{4.3}
\end{equation}
Equality in (\ref{4.3}) serves an estimate for $Q_{\rm SOC}$ and yields
$Q_{\rm SOC} \simeq 60$nW/cm$^2$ under Earth's gravity, in very reasonable 
agreement with the experimental result.

The question now remains as to the nature of the SOC state
below $T_\lambda$.  The state must undergo some kind of
transition to superfluidity, but the fact that it continues to support 
a finite temperature gradient appears inconsistent with the isothermal 
nature of a superfluid.  The resolution of this paradox is shown in 
Fig.~\ref{fig5}.  Let $T_\infty(z) = T_\lambda(z) - \Delta T_\infty(Q)$ and 
$T_c(z) = T_\lambda(z) - \Delta T_c(Q)$ define local values of the interface 
and instability temperatures discussed in Sec.\ III.B, where $\Delta T_c 
= T_{\lambda,0} - T_c(Q)$ and $\Delta T_\infty = T_{\lambda,0} 
- T_\infty(Q)$ are the deviations from $T_\lambda$ in zero gravity.
For $g=0$ an interface state, represented by the upper 
curve in Fig.~\ref{fig5}, is formed.  For $g > 0$, $T(z)$ 
first drops below the local transition at a point $z_0$, and attempts 
to asymptote to an isothermal superfluid at a temperature close to 
$T_\infty(z_0)$.  However, at a point $z_1 > z_0$, the descending 
line $T_c(z_1)$ meets $T_\infty(z_0)$ and the superfluid becomes unstable.
A vortex is generated and crosses the cell, leading to dissipation and 
a finite temperature drop across it.  By this mechanism the temperature is 
able to drop below the instability temperature and asymptote once again to
an isothermal superfluid at a temperature close to $T_\infty(z_1)$.
The entire scenario then repeats itself approximately periodically 
at a sequence of points $z_n$, $n=1,2,3,\ldots$, where $T_c(z_n)$ meets
$T_\infty(z_{n-1})$.  The result is a dynamic staircase structure,
a snapshot of which is represented by the lower curve in Fig.~\ref{fig5}.
This structure fluctuates in time as vortices form and annihilate, but
is found numerically to slowly move in an escalator-like fashion upstream 
\cite{WM00}.

The time resolution of present thermometry is far too poor to detect 
fluctuations in the temperature profile, whose mean would correspond to 
a straight line parallel to and somewhere between $T_\infty(z)$ and 
$T_c(z)$.\footnote{A renormalization group calculation of this mean profile 
as a function of $Q$ in the large-$N$ limit is described in \cite{H99a,H99b}.} 
Experiments to detect the predicted stream of vortices, via the noise they 
emit in the form of second sound, are in the planning stages.  Note that 
only if further numerical or experimental investigation were to reveal power 
law spatial and/or temporal correlations in the vortex nucleation events 
(as exhibited, for example, in $1/f$-type second sound noise spectra) 
would one obtain \emph{a postiori} evidence in support of the ``C'' 
in SOC.

A simple calculation shows that the distances between vortices must scale 
as $z_n-z_{n-1} \approx [\Delta T_\infty(Q) - \Delta T_c(Q)] g_e/\gamma g$, 
increasing with larger $Q$ and smaller $g$.  Although the SOC state requires 
finite gravity, the regime of widely separated vortices may in the future 
prove sufficiently interesting that a controlled low gravity experiment 
will become desirable.

\section{Specific heat at constant heat current} 

\subsection{Enhanced specific heat}

The presence of a heat current is predicted to enhance the specific heat 
of superfluid $^4$He above its equilibrium value, $C_0$.  This can easily
be seen at low $Q$, sufficiently far below $T_\lambda$, where the free 
energy enhancements are $\Delta F(T,{\bf U}_s) = {1\over 2} \rho_s 
{\bf U}_s^2$ and $\Delta \Phi(T,{\bf J}_s) = -{\bf J}_s^2/2 \rho_s$, 
in which $\rho_s(T) \approx \rho_0 |\epsilon|^\zeta$ is the equilibrium 
superfluid density.  Thus, at fixed superfluid velocity ${\bf U}_s$,
\begin{equation} 
\Delta C_{U_s} \equiv C_{U_s} - C_0 \approx T \rho_s {\bf U}_s^2 \zeta 
(1-\zeta)/2 T_\lambda^2 |\epsilon|^2 > 0,
\label{5.1}
\end{equation}
while at fixed heat current ${\bf Q}$, 
\begin{equation}
\Delta C_Q \equiv C_Q - C_0 \approx T {\bf J}_s^2 \zeta (1+\zeta)/2 
T_\lambda^2 \rho_s |\epsilon|^2 > 0.
\label{5.2}
\end{equation}  

Closer to $T_\lambda$ the superfluid density depends strongly on heat 
current and the apparent divergences at $T_\lambda$ in (\ref{5.1})
and (\ref{5.2}) are replaced by new singularities at the phase boundary 
$T_c(U_s)$ shown in Fig.~\ref{fig1}(b).  It is predicted that $\Delta 
C_{U_s}$ will remain finite, rising to a cusp at the phase boundary 
\cite{HD94}, while $\Delta C_Q$ is predicted to diverge \cite{cq96,HD96}.  
The latter result follows on very general grounds from the thermodynamic 
conjugacy of ${\bf U}_s$ and ${\bf J}_s$ discussed in Sec.\ II.  The 
usual thermodynamic manipulations imply the relation    
\begin{equation}
C_Q = C_{U_s} + T \left({\partial J_s \over{\partial T}}\right)^2_{U_s}
\left({\partial U_s \over{\partial J_s}}\right)_T. 
\label{5.3}
\end{equation}
Since the susceptibility, $\left(\partial U_s /\partial J_s \right)_T$
[proportional to the slope of the curve in Fig.~\ref{fig2}(b)] is
expected to diverge at the phase boundary, while $(\partial J_s/\partial T)_{U_s}
= U_s (\partial \rho_s/\partial T)_{U_s}$ remains finite, $C_Q$ will exhibit
a \emph{divergent} enhancement.

There have been no measurements of $C_{U_s}$ to date, but an experiment of 
this kind might be performed in the presence of a persistent current 
flowing around a loop, similar to the superfluid gyroscope experiment \cite{CR72},
where in the absence of vortices ${\bf U}_s$ indeed remains fixed as $T$ is 
varied.  A measurement could prove difficult due to the small magnitude of the
enhancement and the challenge of holding ${\bf U}_s$ constant while measuring
the specific heat.  However, it has been suggested that with very fast
thermometry, a measurement of $C_{U_s}$ might be obtained through a
second-sound measurement where the second-sound waves are propagated
perpendicular to ${\bf U}_s$ \cite{H97}.

Measurement of $C_Q$ is more straightforward.  The predicted divergence
of $\Delta C_Q$, together with the fact that ${\bf Q}$ can be experimentally 
controlled with great precision, implies a much more visible experimental 
signature.  The first experimental measurements of this quantity were recently 
reported \cite{cq99}.  As will be discussed below, they indicate that the heat 
capacity is indeed enhanced, but with a magnitude that is significantly larger 
than theoretical predictions.

\subsection{The superfluid breakdown temperature}

Experiments performed at constant $Q$ should find that $C_Q$ diverges at 
a temperature $T_c(Q) < T_\lambda$, defined by inverting (\ref{2.20}):
\begin{equation}
|\epsilon_c(Q)| = \frac{T_\lambda
-T_c(Q)}{T_\lambda} = \left( \frac{Q}{Q_0^c} \right)^{x}, 
\label{5.5} 
\end{equation}
where $Q_0^c = Q_0 y_c$ and $x = 1/\Delta_Q = 1/2\nu \simeq 0.746$
\cite{O84,HD91,GCH96}.  Based on a renormalization group analysis of 
the Model F equations \cite{HH77}, in an approximation neglecting 
vortices (and hence decay of superflow), the prediction $Q_0^c 
\approx 7.4$~kW/cm$^2$ was obtained \cite{HD92b}.  More recently, 
using an extension of this theory, accounting for dissipation
within a large-$N$ approximation, the value $Q_0^c \approx 6.6$~kW/cm$^2$ 
was obtained \cite{H99b}.

\begin{figure}
\centerline{\psfig{figure=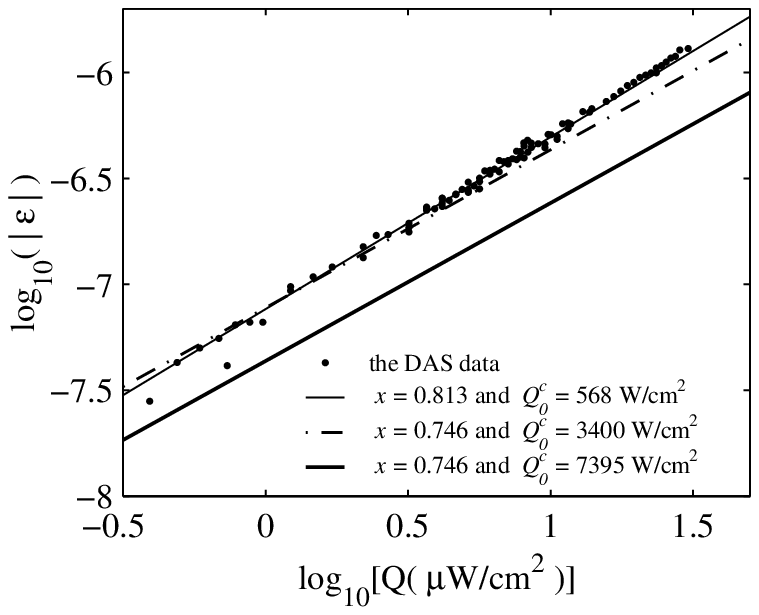,width=3.4in}}
\vspace{0.1in}
\caption{Thick solid line:  $T_c(Q)$, the theoretically predicted 
temperature of superfluid breakdown \protect\cite{HD92b}.  Thin solid 
line:  $T_{\rm DAS}(Q)$, where the exponent $x$ and amplitude $Q_0^c$ are
chosen to best fit to the observed temperature of superfluid breakdown 
represented by the data points \protect\cite{DAS88}.  Dashed-dotted line: 
the value of $Q_0^c$ that, along with the theoretical value $x=1/2\nu$, 
makes the experimental heat capacity data match the more recent theoretical 
prediction for the scaling function \protect\cite{H99b} (see 
Fig.~\protect\ref{fig10} below).}
\label{fig6}
\end{figure}

These theoretical results disagree with the results of thermal conductivity 
experiments \cite{DAS88}.  The onset of thermal resistance was found to occur 
at a temperature which we call $T_{\rm DAS}(Q)$ that obeys (\ref{5.5}), but 
with $x = 0.813 \pm 0.012$ and $Q_0^c = 568 \pm 200$~W/cm$^2$.  This is a
curve in the $T$-$Q$ plane that falls below the theoretically estimated 
$T_c(Q)$ for all experimentally accessible temperatures (see Fig.~\ref{fig6}).  
A number of explanations have been proposed for the discrepancy:  (a) the 
transition at $T_{DAS}(Q)$ may be caused by a temperature instability at the 
cell wall associated with the singular Kapitza resistance [which raises the
temperature near the bottom (heated) end plate above that of the bulk, which
therefore could serve as a vortex nucleation center], and hence lies below $T_c(Q)$ 
\cite{cq99}; (b) $T_{\rm DAS}$ may be related to a gravity-dependent transition, 
again lying below $T_c(Q)$, and will increase towards $T_c(Q)$ as gravity is 
reduced, e.g., by going into space \cite{H99b}; (c) since the transition is 
only sharply defined in the absence of vortices, $T_{\rm DAS}(Q)$ might be 
analogous to a spinodal line in a first-order phase transition \cite{LA96}:  
fluctuation-induced vortices nucleate the transition to the dissipative 
phase, and $T_{\rm DAS}(Q)$ will differ from experiment to experiment, 
depending on the heating rate used.  Since any superfluid state above 
$T_\infty(Q)$ should be unstable by this mechanism, an infinitessimally 
slow experiment should find the transition at $T_\infty(Q)$.

It is possible that all of these effects (and perhaps others) are present.  
The real question, whose resolution clearly requires more experimental data, 
is which one imposes the most severe limitation on present experiments.
The answer to this question has implications for the measurement of $C_Q(T)$.  
If effect (c) is dominant, then experiments have basically already reached 
the intrinsic limit on how close they can approach the divergence of $C_Q(T)$
(for the range of $Q$ explored thus far), though it may be possible to design
an experiment with faster heating rates and fast enough thermometers to reach
$T_c(Q)$ before a vortex can nucleate.  On the other hand, if proposal (b) 
is correct, a space-based microgravity measurement of $C_Q(T)$ should be able 
to get considerably closer to $T_c(Q)$ than one performed on the ground.  If 
proposal (a) is correct, carefully designed ground-based experiments 
might be able to approach $T_c(Q)$ more closely:  if $T_{\rm DAS}$ is due to
a boundary effect, a cell constructed with a bottom plate that is much larger 
than the cross-sectional area of the bulk helium sample could decrease the
singular Kapitza resistance sufficiently so that the bulk helium can reach
$T_c(Q)$ without a boundary instability interfering.  A cell of this 
configuration that maintains a reasonable geometry for heat flow would 
have to be fairly tall and might therefore be more susceptible to gravity
effects.

\begin{figure}[t] 
\centerline{\psfig{figure=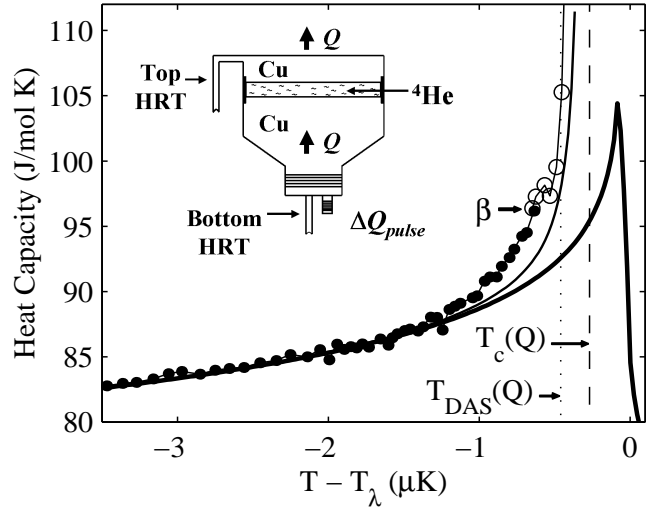,width=3.4in}}
\vspace{0.1in} 
\caption{Sample heat capacity data for $Q=3.5~\mu {\rm W/cm}^2$ 
\protect\cite{cq99}.  Thick solid line:  equilibrium $C_0$ obtained from 
a fit to the Lambda Point Experiment (LPE) data \protect\cite{lpe96} that is 
subsequently rounded for gravity.  Thin solid line:  theoretical prediction 
\protect\cite{H99b} that includes vortices (rounded for gravity).  Solid 
circles:  data from the average of the top and bottom thermometers.  Open 
circles:  data from the top thermometer only [beyond the point marked $\beta$, 
where the bottom (hotter) thermometer was found to change its behavior, perhaps 
as a result of a boundary heating effect].  The temperature $T_{\rm DAS}$ marks 
the onset of dissipation found in earlier thermal conductivity experiments 
\protect\cite{DAS88}, while $T_c(Q)$ is estimated from a certain theoretical 
fit to the data discussed later in the text.  Inset:  schematic diagram of 
the experimental cell.  HRT stands for high resolution thermometer.} 
\label{fig8} 
\end{figure}

\subsection{Experimental measurements}

The first experimental measurements of $C_Q(T)$ \cite{cq99} confirm 
the predicted enhancement, but find that its magnitude is significantly
larger than current predictions \cite{cq96,HD96,H99b}.
The data were taken over the range $1~\mu {\rm W/cm}^2 \leq Q \leq 
4~\mu {\rm W/cm}^2$, and a representative set at $Q=3.5~\mu {\rm W/cm}^2$ 
is shown in Fig.~\ref{fig8}.

\begin{figure}[t]
\centerline{\psfig{figure=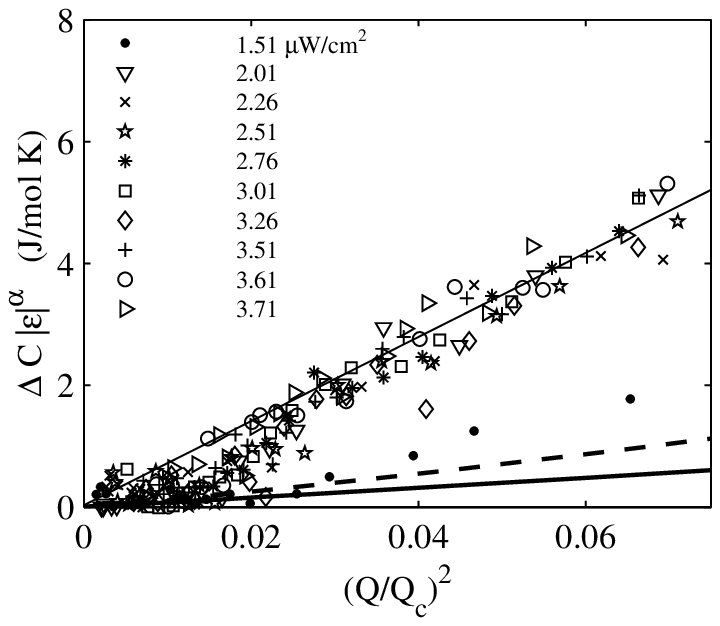,width=3.4in}}
\vspace{0.1in}
\caption{Scaling plot of the differential heat capacity measurements for 
various values of $Q$ \protect\cite{cq99}.  The experimental data are scaled 
using $Q_0^c = 6.6~{\rm kW/cm}^2$, and terminated at the temperature indicated 
by $\beta$ in Fig. \protect\ref{fig8}.  The predicted collapse of the
data for different $Q$ values onto nearly the same curve verifies the
basic scaling hypothesis (\protect\ref{5.6}).  Thin solid line:  straight 
line fit to the data.  Thick solid line:  theoretical prediction neglecting 
dissipation \protect\cite{cq96,HD96}.  Dashed line:  theoretical prediction 
including dissipation \protect\cite{H99b}.  Neither theoretical curve is rounded 
for gravity.}
\label{fig9}
\end{figure}

The extent of the disagreement between theory and experiment can be observed 
more clearly when the data are plotted in scaled form.  All theories predict 
that the enhancement should obey a scaling form
\begin{equation}
\Delta C_Q = |\epsilon|^{-\alpha} f_{J_s}\left[{Q \over Q_c(T)} \right],
\label{5.6}
\end{equation}
where $f_{J_s}(x)$ is a universal scaling function.  From (\ref{5.2})
and the scaling relations in Table~\ref{table1}, for $x \ll 1$ one has
\begin{eqnarray}
f_{J_s}(x) &=& f_2 x^2 + O(x^4) \nonumber \\
f_2 &\equiv& [\zeta (\zeta + 1) (Q_0^c)^2/2\rho_0 T_\lambda^3 S_\lambda^2].
\label{5.7}
\end{eqnarray}
Using the theoretical estimates one obtains $f_2 = 8.9$~J/mol~K 
for $Q_0^c = 7.4$~kW/cm$^2$ and $f_2 = 7.0$~J/mol~K for $Q_0^c = 6.6$~kW/cm$^2$,
where the molar volume 27.38~cm$^3$/mol has been used to obtain familiar units.

In Fig.~\ref{fig9} we show the heat capacity enhancement as a function of
the scaling variable $(Q/Q_c)^2$.  Since $Q_c(T)$ is not actually measured
in the experiment, we scale the data using the theoretically predicted form 
$Q_c = Q_0^c |\epsilon|^{\Delta_Q}$ with $Q_0^c = 6.6~{\rm kW/cm}^2$ and
$\Delta_Q = 2\nu = 1.342$ \cite{H99b}.  As anticipated, the data for all
$Q \ge 2~\mu {\rm W/cm}^2$ collapse onto a single linear curve, verifying
that the exponent $\Delta_Q$ is at least consistent with the data.  However, 
the slope of the experimental line is $f_2^{\rm expt} = 69 \pm 4$~J/mol K,
approximately ten times larger than the theoretical prediction.

\begin{figure}[t] 
\centerline{\psfig{figure=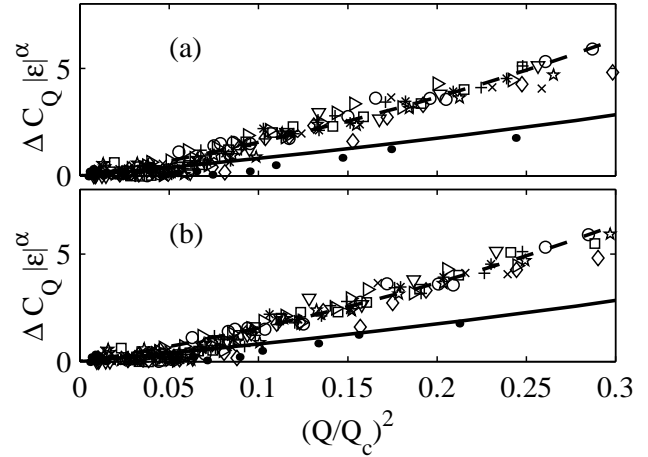,width=3.4in}}
\vspace{0.1in} 
\caption{Alternative scaling plot of the differential heat capacity measurements 
for various values of $Q$ \protect\cite{cq99}.  The data symbols are the same as 
those used in Fig.~\protect\ref{fig9}.  The experimental data are scaled using
$Q_c(T)$ derived from (\ref{5.5}) by (a) using the theoretical exponent value 
$x = 1/2\nu = 0.746$, but amplitude $Q_0^c = 3.4~{\rm kW/cm}^2$ chosen to best 
match the theoretical scaling function; and by (b) assuming that $T_c(Q) \approx 
T_{DAS}(Q)$ with $Q_0^c = 0.65~{\rm W/cm}^2$ and $x = 0.813$.  For the $Q$-range 
of the experimental data, the two analyses are basically identical, with the lines 
completely overlapping to well within experimental resolution.  Only for higher 
$Q$ data would the two fits become distinguishable.  Solid line: theoretical 
prediction neglecting dissipation \protect\cite{cq96,HD96};  Dashed line:  
theoretical prediction including dissipation \protect\cite{H99b}.}
\label{fig10} 
\end{figure}

An optimistic explanation for the discrepancy between theory and
experiment is that the theories are producing reasonable estimates for
the universal scaling function $f_{J_s}(x)$, but that the nonuniversal 
amplitude $Q_0^c$, depending on detailed properties of the $^4$He
system and therefore more difficult to compute, is estimated 
less accurately.  As illustrated in Fig.~\ref{fig10}(a), the choice $Q_0^c 
= 3.4$~kW/cm$^2$\footnote{This value is in fact within the estimated margin 
of error for the amplitude calculation.  The uncertainty of the theory is 
a factor $< 2$ (R. Haussmann, private communication), while the adjustment 
here is $\simeq 1.9$.} in fact places the experimental data on top 
of the more recent theoretical curve \cite{H99b}.  With this choice, $T_c(Q)$ 
lies somewhat above $T_{\rm DAS}(Q)$ (vertical dashed line in Fig.~\ref{fig6}).
A somewhat smaller choice for $Q_0^c$ would provide an equally good fit to
the earlier theory \cite{HD92b}.  Unfortunately, all of the data lie at 
fairly small $(Q/Q_c)^2 \leq 0.3$ where the scaling function has little 
structure (essentially indistinguishable from linear within the scatter 
of the data).  A true test would require data in the regime $(Q/Q_c)^2 
\to 1$ where the scaling function diverges.

For completeness we also show in Fig.~\ref{fig10}(b) an equally good 
scaling collapse based on the assumption that $T_c(Q) \approx T_{\rm DAS}(Q)$.  
Thus, we use $Q_c(T)$ derived from (\ref{5.5}) using $x=0.813$ (i.e., 
effectively $\nu = 0.615$) and $Q_0^c = 0.65$~kW/cm$^2$ [optimally chosen
within the error bars quoted in \cite{DAS88}].  Since $T_{\rm DAS}$ places 
a lower bound on $T_c(Q)$, the sharpest conclusion we can make at this stage 
is that the data are consistent with the scaling hypothesis for a fairly broad 
range of experimentally and theoretically motivated parameter choices and that 
more data closer to $T_c(Q)$ will be required for a critical test of the theory. 

The experimental measurements were taken in a cell that was only 0.64~mm
high, about as close to the optimal height as practical considerations allow. 
Although the effects of gravitational rounding were therefore minimized, they 
were not entirely eliminated.  The variation of $T_\lambda$ across the cell due 
to gravity was $\delta T_\lambda \simeq 8 \times 10^{-8}$~K.  A reasonable
criterion for a data point to be unaffected by gravity is that one should
have $|T-T_\lambda| \ge 10 \delta T_\lambda$.  This would restrict the 
temperature range of the experiment to more than about $1 \mu$K below 
$T_\lambda$.  Essentially none of the interesting data shown in 
Figs.~\ref{fig8}--\ref{fig10} satisfy this criterion.  In fact, because 
measurements \cite{BAKF99} have shown that under a heat flux $Q > 4~\mu 
{\rm W/cm}^2$, helium exhibits appreciable dissipation, there is no range 
of parameters for which this criterion can be satisfied under Earth's 
gravity in an isothermal experiment that measures $C_Q$ close to $T_c(Q)$. 

The small cell height used in the experiment precluded the use of
side-wall thermometry.  The helium temperature was measured using
thermometers mounted on the cell end-plates and the data were corrected
for the singular Kapitza resistance \cite{FBKA}. The temperature range
of the measurements was limited because the bottom (hotter) thermometer
changed its behavior before the bulk helium temperature reached
$T_{DAS}(Q)$ \cite{cq99}. It was proposed that this change was due to
another boundary effect related to the Kapitza resistance.  The
temperature at which this phenomenon occurred is indicated by $\beta$ in
Fig.~\ref{fig8}, and is the maximum temperature of the data shown in
Figs.~\ref{fig9} and \ref{fig10}.  As a result of the reduced range, the
data in Fig.~\ref{fig10} do not reach high enough temperatures to
encounter much curvature in the scaling function.  Measurements that
approach closer to the divergence are clearly needed.  Data up to
$T_{DAS}(Q)$ should be easily obtainable using a deeper cell constructed
with a mid-plane thermometer. However, since rounding due to gravity
will be even more of a detriment over the region where the scaling
function has significant curvature, the ill-effects of the
deeper cell height will obscure some of the benefit gained by the
increased temperature range. 

The only definitive way to circumvent the problems raised in the
previous two paragraphs is to perform a space-based measurement of
$C_Q$.  An experiment in the absence of gravity would obtain data up to
$T_{\rm DAS}(Q)$ without gravitational rounding, permitting an extension 
of the scaling data into a more revealing temperature range.  This would 
allow a considerably improved estimate of where $T_c(Q)$ lies in relation 
to $T_{\rm DAS}(Q)$ [in particular if $\Delta C_Q$ is still finite at 
$T_{\rm DAS}(Q)$ one would conclude that $T_c(Q) > T_{\rm DAS}(Q)$].  
Furthermore, a space experiment would test the suggestion that 
$T_{\rm DAS}(Q)$ is a gravity artifact \cite{H99b}, and permit 
the construction of a deep cell with asymmetric endplates to test 
whether $T_{\rm DAS}(Q)$ is a boundary effect.

\section{Conclusions}

We have reviewed the current status of experiment and theory of liquid
helium when it is very close to the superfluid transition, but not in 
equilibruim.  Nonequilibruim states are most easily probed experimentally 
by passing a uniform heat flux, ${\bf Q}$, through the liquid.  If the 
average temperature of the sample is sufficiently far above $T_\lambda$, 
the heat flux produces a uniform temperature gradient, given by (\ref{3.1}).  
Below $T_\lambda$, a sufficiently small $Q$ produces a uniform temperature.  
Between these limits there is a nonlinear regime that is only partially 
understood.

In the uniform temperature regime below $T_\lambda$, ${\bf J}_s$ 
(proportional to ${\bf Q}$) and ${\bf U}_s$ may be treated as conjugate 
thermodynamic variables [see (\ref{2.4})].  By analogy to equilibrium 
critical point phase transitions, one can imagine a critical region in 
the $Q$-$T$ plane, and we are able to define new critical point exponents 
for these conjugate variables, and derive scaling laws between them that 
permit all of them to be evaluated in the critical region.  These 
predictions have undergone only limited experimental tests.

In all instances, the nature of the nonequilibrium phase transition 
is strongly affected by gravity. If the heat flux is in the opposite
direction to gravity, the transition can be made to occur within the
cell, the bottom end where the heat enters being normal, and the top,
where the heat is removed, superfluid.  In this case, the width of the
interface, which is the nonlinear region, is dominated by Earth's 
gravity at the very low values of $Q$ that are of primary interest
to investigations of critical phenomena, and is too small to be probed 
experimentally.  If the heat flux is in the same direction as gravity, 
the system develops a temperature gradient equal to the gravity induced 
gradient in $T_\lambda$, so that it has the same reduced temperature 
everywhere.  This is the so-called SOC state.  At small $Q$, the SOC 
state can be on the normal side of the transition.  At larger $Q$ it 
exists on the superfluid side, the temperature gradient being maintained 
by a dynamic mechanism of vortex production.

On the superfluid side of the transition, the heat capacity, $C_Q$, is
expected to be larger than $C_Q(Q=0)$, and in fact to diverge at the 
phase boundary.  Preliminary experiments show that $C_Q$ is indeed enhanced,
and in fact is far larger than expected theoretically.  It is not known
whether this discrepancy is due to uncertainty in the position of the
phase boundary or to other causes.  This result too is strongly affected
by gravity, which introduces an imhomogeneity in the reduced temperature,
$\epsilon$, at the small values of $\epsilon$ necessary for these
experiments.

Although much progress has been made in understanding this nonequilibrium
phase transition and exploring the exciting new physics that it exhibits, 
it is clear that experiments in the absence of Earth's gravity will be 
needed.  One such expriment (DYNAMX) is in preparation, and others are 
planned.


\begin{references}

\harvarditem{Ahlers and Liu}{1996}{AL96} Ahlers, G., and F.-C. Liu, 1996,
J. Low Temp.\ Phys.\ {\bf 105}, 255.

\harvarditem{Baddar {\it et al.}}{1999}{BAKF99} Baddar, H., G. Ahlers, 
K. Kuehn, and H. Fu, 1999, preprint.

\harvarditem{Bak {\it et al.}}{1988}{BTW88} Bak, P., C. Tang and K. Wiesenfeld,
1988, Phys.\ Rev.\ A {\bf 38}, 364.

\harvarditem{Chui {\it et al.}}{1996}{cq96} Chui, T. C. P., D. L. Goodstein,
A. W. Harter and R. Mukhopadhyay, 1996, Phys.\ Rev.\ Lett.\ {\bf 77}, 1793.

\harvarditem{Chui {\it et al.}}{1997}{DXDF27} Chui, T. C. P., P. B. Weichman,
R. Galletly, D. Elliott, K. Aaron and D. L. Goodstein, 1997, {\it DYNAMX vibration 
requirement}, DYNAMX Design File \#27 (DX-DF-27).

\author{DYNAMX vibration splinter group: \\ Talso Chui, Peter Weichman, 
Bob Galletly, Dave Elliott, \\ Kim Aaron, and David Goodstein}

\harvarditem{Clow and Reppy}{1972}{CR72} Clow, J. R., J. D. Reppy, 
1972, Phys.\ Rev.\ A {\bf 5} 424.

\harvarditem{Day {\it et al.}}{1998}{unm98} Day, P. K., W. A. Moeur, 
S. S. McCready, D. A. Sergatskov, F.-C. Liu and R. V. Duncan, 1998, 
Phys.\ Rev.\ Lett.\ {\bf 81}, 2474.

\harvarditem{Dingus, {\it et al.}}{1986}{DZM86} Dingus, M., F. Zhong and
H. Meyer, 1986, J. Low Temp.\ Phys.\ {\bf 65}, 185.

\harvarditem{Dohm}{1991}{D91} Dohm, V., 1991, Phys.\ Rev.\ B {\bf 44},
2697 (1991).

\harvarditem{Duncan {\it et al.}}{1988}{DAS88} Duncan, R. V., G. Ahlers 
and V. Steinberg, 1988, Phys.\ Rev.\ Lett.\ {\bf 60}, 1522.

\harvarditem{Duncan and Ahlers}{1991}{DA91} Duncan, R. V. and G. Ahlers, 
1991, Phys.\ Rev.\ B {\bf 43}, 7707.

\harvarditem{Fisher}{1973}{F73} Fisher, M. E., 1973, Proc.\ Nobel Symp.\ 
{\bf 24}, 16.

\harvarditem{Fisher}{1983}{F83} Fisher, M. E., 1983, {\it Scaling, universality
and renormalization group theory}  in Critical Phenomena, Lecture Notes in
Physics (Springer-Verlag, Berlin), {\bf 186}, 1,  F. J. W. Hahn, editor.

\harvarditem{Fu {\it et al.}}{1998}{FBKA} Fu, H., H. Baddar, K. Kuehn, and 
G. Ahlers, 1998, Low Temp.\ Phys.\ {\bf 24}, 69.

\harvarditem{Giordano {\it et al.}}{1987}{GMB87}  Giordano, N., P. Muzikar, 
and S. S. C. Burnett, 1987, Phys.\ Rev.\ B {\bf 36}, 667.

\harvarditem{Goldner and Ahlers}{1992}{GA92} Goldner, L., and G.~Ahlers, 
1992, Phys.\ Rev.\ B {\bf 45}, 13129.

\harvarditem{Goodstein {\it et al.}}{1996}{GCH96} Goodstein, D. L., T. C. P.
Chui, and A. W. Harter, 1996, Phys.\ Rev.\ Lett.\ {\bf 77}, 979.

\harvarditem{Greywall and Ahlers}{1973}{GA73} Greywall, D.S., and G. Ahlers,
1973, Phys.\ Rev.\ A {\bf 7}, 2145.

\harvarditem{Harter {\it et al.}}{2000}{cq99} Harter, A. W., R. A. M. Lee,
A. Chatto, X. Wu, T. C. P. Chui and D. L. Goodstein, 2000, 
Phys.\ Rev.\ Lett.\ {\bf 84}, 2195.

\harvarditem{Haussmann}{1997}{H97} Haussmann, R., 1997, J. Low Temp.\ 
Phys.\ {\bf 107}, 21.

\harvarditem{Haussmann}{1999a}{H99a} Haussmann, R., 1999a, J. Low Temp.\ 
Phys.\ {\bf 114}, 1.

\harvarditem{Haussmann}{1999b}{H99b} Haussmann, R., 1999b, Phys.\ Rev.\ B 
{\bf 60}, 12349.

\harvarditem{Haussmann and Dohm}{1991}{HD91} Haussmann, R., and V. Dohm, 
1991, Phys.\ Rev.\ Lett.\ {\bf 67}, 3404.

\harvarditem{Haussmann and Dohm}{1992a}{HD92} Haussmann, R., and V. Dohm, 
1992a, Z. Phys.\ B {\bf 87}, 229.

\harvarditem{Haussmann and Dohm}{1992b}{HD92b} Haussmann, R., and V. Dohm, 
1992b, Phys.\ Rev.\ B {\bf 46}, 46.

\harvarditem{Haussmann and Dohm}{1994}{HD94} Haussmann, R., and V. Dohm, 
1994, Phys.\ Rev.\ Lett.\ {\bf 72}, 3060.

\harvarditem{Haussmann and Dohm}{1996}{HD96} Haussmann, R., and V. Dohm, 
1996, Czech.\ J. Phys.\ {\bf 46-S1}, 171.

\harvarditem{Hohenberg {\it et al.}}{1976}{HAHS76} Hohenberg, P. C., 
A. Aharony, B. I. Halperin and E. D. Siggia, 1976, Phys.\ Rev.\ B {\bf 13},
2986.

\harvarditem{Hohenberg and Halperin}{1977}{HH77} Hohenberg, P. C., 
and B. I. Halperin, 1977, Rev.\ Mod.\ Phys.\ {\bf 49}, 435.

\harvarditem{Hohenberg and Martin}{1965}{HM65} Hohenberg, P. C., 
and P. C. Martin, 1965, Ann.\ Phys.\ (N.Y.) {\bf 34}, 291.

\harvarditem{Landau}{1941}{L41} Landau, L.D. 1941, USSR {\bf 5}, 71. 
[English translation: 1965, {\it Collected Papers of L.D. Landau}, D. ter
Haar, editor, (Gordon and Breach, New York), p.\ 301.]

\harvarditem{Langer and Fisher}{1967}{LF67} Langer, J. A., and M. E. 
Fisher, 1960, Phys.\ Rev.\ Lett.\ {\bf 19}, 560 (1960).

\harvarditem{Lipa {\it et al.}}{1996}{lpe96} Lipa, J. A., D. R. Swansson,
J. A. Nissen, T. C. P. Chui and U. E. Israelson, 1996, Phys.\ Rev.\ Lett.\
{\bf 76}, 944.

\harvarditem{Liu and Ahlers}{1996}{LA96} Liu, F.-C., and G. Ahlers, 
1996, Phys.\ Rev.\ Lett.\ {\bf 76}, 1300.

\harvarditem{Machta {\it et al.}}{1993}{MCH93} Machta, J., D. Candela, 
and R. B. Hallock, 1993, Phys.\ Rev.\ B {\bf 47}, 4581.

\harvarditem{Moeur {\it et al.}}{1997}{soc97} Moeur, W. A., 
P. K. Day, F.-C. Liu, S. T. P. Boyd, M. J. Adriaans, and R. V. Duncan, 
1997, Phys.\ Rev.\ Lett.\ {\bf 78}, 2421.

\harvarditem{Muzikar and Giordano}{1989}{MG89} Muzikar, P., and
N. Giordano, 1989, Physica A {\bf 157}, 742.

\harvarditem{Onuki}{1983}{O83} Onuki, A., 1983, J. Low Temp.\ Phys.\ 
{\bf 50}, 433.

\harvarditem{Onuki}{1984}{O84} Onuki, A., 1984, J. Low Temp.\ Phys.\
{\bf 55}, 309.

\harvarditem{Onuki}{1987}{O87} Onuki, A., 1987, Jpn.\ J. Appl.\ Phys.\
{\bf 26}, 365.

\harvarditem{Onuki}{1996}{O96} Onuki, A., 1996, J. Low Temp.\ Phys.\ 
{\bf 104}, 133.

\harvarditem{Onuki and Yamazaki}{1996}{OY96} Onuki, A., and Y. Yamazaki, 
1996, J. Low Temp.\ Phys.\ {\bf 103}, 131.

\harvarditem{Pfeuty {\it et al.}}{1974}{PJF74} Pfeuty, P., D. Jasnow
and M. E. Fisher, 1974, Phys.\ Rev.\ B {\bf 10}, 2088.

\harvarditem{Tam and Ahlers}{1985}{TA85} Tam, W. Y., and G. Ahlers,
1985, Phys.\ Rev.\ B {\bf 32}, 5932; {\bf 33}, 183 (1986).

\harvarditem{Weichman and Miller}{2000}{WM00} Weichman, P. B., and J. Miller,
2000, J. Low Temp.\ Phys.\ {\bf 119}, 155.

\harvarditem{Weichman {\it et al.}}{1998}{WPMM98} Weichman, P. B., A. Prasad, 
R. Mukhopadhyay and J. Miller, 1998, Phys.\ Rev.\ Lett.\ {\bf 80}, 4923.

\end{references}
\end{document}